# An Empirical study on LLM-based Log Retrieval for Software Engineering Metadata Management

Simin Sun
simin.sun@gu.se
Chalmers University of Technology
and University of Gothenburg
Gothenburg, Sweden

Yuchuan Jin
yuchuan.jin@zenseact.com
Zenseact
Gothenburg, Sweden

Miroslaw Staron
miroslaw.staron@gu.se
Chalmers University of Technology
and University of Gothenburg
Gothenburg, Sweden

## Abstract

Developing autonomous driving systems (ADSs) involves generating and storing extensive log data from test drives, which is essential for verification, research, and simulation. However, these high-frequency logs, recorded over varying durations, pose challenges for developers attempting to locate specific driving scenarios. This difficulty arises due to the wide range of signals representing various vehicle components and driving conditions, as well as unfamiliarity of some developers' with the detailed meaning of these signals. Traditional SQL-based querying exacerbates this challenge by demanding both domain expertise and database knowledge, often yielding results that are difficult to verify for accuracy.

This paper introduces a Large Language Model (LLM)-supported approach that combines signal log data with video recordings from test drives, enabling natural language based scenario searches while reducing the need for specialized knowledge. By leveraging scenario distance graphs and relative gap indicators, it provides quantifiable metrics to evaluate the reliability of query results. The method is implemented as an API for efficient database querying and retrieval of relevant records, paired with video frames for intuitive visualization. Evaluation on an open industrial dataset demonstrates improved efficiency and reliability in scenario retrieval, eliminating dependency on a single data source and conventional SQL.

## 1 Introduction

Autonomous driving software is complex in terms of functionality and testing. The software must be capable of real-time processing of several signals from sensors [2], create a model of the driving situation, and then send signals to the actuators, preserving the safety of the passengers. Testing this software is even more challenging, requiring traditional methods like unit and system testing, as well as real-world and simulation testing for autonomous vehicles [13]. Thorough testing also involves collecting rare scenario data and annotating existing data [16]. Machine learning (ML) patterns, such as Data Lakes, are already employed to collect data from test vehicles [14]. Test vehicles are equipped with numerous sensors and LiDAR systems to test and assess new features in on-road testing. During these tests, sensor metadata is processed by onboard software, generating actuator signals. This signal data is logged in files for developers and testers to analyze and optimize vehicle safety, comfort, and performance. However, different components record data at varying frequencies, resulting in approximately 100 records per second. This generates a massive volume of data stored in databases and makes it difficult to locate specific scenarios.

In current software development practice, developers rely on SQL to query specific scenarios. The challenge arises when they attempt to craft queries for complex conditions. For instance, a developer needs to find a scenario such as "driving the vehicle in the rain at a speed of 45 km per hour". The sheer volume of signal types, coupled with the complexity of understanding each signal's purpose and use, makes it nearly impossible for any developer to construct effective queries without extensive expertise. Secondly, signal names are long, sometimes exceeding 150 characters (e.g., \*\*\*\*\*_qm_sensorfusion_a/\*\*\*\*_qm_sensorfusion_a_vehicle _motion_stat e_data/data/lateral_velocity/velocity/me-ters_per_second/value). The length and complexity of the signal names translate to the length and complexity of the columns and tables in the databases where the signal data are stored. It becomes challenging to recall the exact header for querying, especially when multiple signals are related to the same keyword and multiple keywords appear in the scenario. Crafting SQL queries requires engineers to understand not only the signals, but also the database structure.

Crafting the SQL query is one challenge, and another is the evaluation process. The lack of labels complicates evaluation, as creating labels is difficult due to the fact that records can be described from multiple perspectives. As a result, the output can only be assessed by checking its general alignment with the query, rather than identifying an exact or optimal match. This means that after a query is executed, the results must be manually evaluated, as there is no quantitative metric to measure the accuracy or relevance of the retrieved records.

In this paper, we focus on enhancing the usability of log data in autonomous driving by allowing software engineers to use natural language queries to retrieve relevant driving scenarios. We address the following research questions:

- **RQ1:** How do different data modalities, models and prompts affect description generation?
- **RQ2:** How do query difficulty and the keyword selection impact the quality and effectiveness of data retrieval?
- **RQ3:** Which metrics and indicators can be used to assess the query results?

To address these questions, we are conducting an empirical study with our industrial partner, Zenseact. To illustrate the results, in this paper, we use an open dataset for replicability, the Zenseact Open Dataset (ZOD) [3], which contains 1,473 publicly accessible sequences. Lastly, we fully utilize video data to enhance scenario descriptions, incorporating motion-related details.

## 2 Related Work

Signal data is generated by processing raw sensor data collected from various equipment using onboard software. Instead of relying on signal data, many studies have explored the direct use of sensor data. Chen et al. [5] labeled LiDAR data from test vehicles for classification tasks, achieving performance comparable to manually labeled data. Baur et al. [4] introduced SLIM, a self-supervised model that distinguishes moving objects from stationary ones. While these models are effective, they are not suited to our research as they mainly provide general classification labels from a single perspective and cannot handle diverse sensor data to generate comprehensive descriptions. Therefore, we used signal data processed by onboard software instead of raw sensor data.

Since the signal data is stored in tables, we explored state-of-the-art methods for querying tabular data. One approach involves training or fine-tuning encoder-based models on labeled table data [11, 21], but this requires labeled data, which we lack. Another option is the decoder-based approach [8, 11, 15], which enhances performance by improving prompts and generated descriptions. However, this method relies on textual information to produce accurate descriptions, and since our data is entirely numeric, generating meaningful descriptions is challenging.

Understanding the context of video data is a popular task in computer vision. Earlier approaches used LSTM-based models [9, 20, 22], focusing on generating captions or summarizing video content. More recent transformer-based models [17, 24, 25] have been developed to generate textual descriptions from video. A growing trend is the use of multi-modal models [10, 12], which leverage multiple data modalities to improve video description generation.

Previous work [7] investigated the use of natural language queries to locate signal log information with the support of video data. However, the approach only utilized a single image frame from each short video sequence, failing to fully exploit the temporal information available in the video. Relying on static images rather than full video omitted critical motion cues—such as distinguishing between moving and stationary states—thereby limiting the expressiveness and accuracy of the generated descriptions. Moreover, this study was based on just 80 records across 10 scenarios, which were specifically chosen to contain events observable in the signal data. While the initial results were promising, the limited size and diversity of the dataset constrained a thorough evaluation of the method's performance and generalizability.

We contribute to our previous work in several key aspects. First, we utilize a different dataset that is openly accessible and approximately 20 times larger than that used in prior studies, significantly enhancing the external validity of our findings. Second, we employ alternative architectures designed to fully leverage the temporal dynamics of video data, rather than relying on a single frame per sequence. Lastly, we scale up the generation process by incorporating a broader set of scenarios that are commonly encountered in real-world practice and span a wide range of difficulty levels.

## 3 Research Design

### 3.1 Data Processing

Zenseact Open Dataset (ZOD) consists of three subsets : Frames, Sequences, and Drives, collected by test vehicles over two years across various European countries. All vehicles were equipped with multiple sensors and cameras. The Frames subset contains annotated camera and LiDAR data, which are outside the scope of our research. Driver consists of 29 multi-minute sequences with full sensor coverage. The volume of this subset is too small for our research. We utilize the subsets – Sequences, which contains 1,473 records of 20-second test-driving sequences. They are in the format of data triplets <ID, signal data, and video data>.

**Signal Data:** The *vehicle_data* file within the sequence dataset corresponds to the signal data. During test drives, sensors generate high-frequency data, which is processed into signals essential for key vehicle systems. All signals are stored in hdf5 format and can be accessed as parquet tables comprising:

- *Vehicle_data:* 19 signals regarding the vehicle status, for example, velocity and acceleration.
- *Vehicle_control_data:* 11 signals of vehicle control status, for example, acceleration pedal, brake pedal.
- *Satellite_data:* 19 signals regarding the satellite data, for example, latitude and longitude.

All three tables will be filtered based on interpretability, with 14 columns being used in this analysis.

**Video Data:** The dataset contains 1,473 video records captured by the front camera, each represented as .jpg images. Each 20-second video is captured at 10 Hz, yielding ten frames per second. An example frame from record #000005 is shown in Figure 1. For each record, we uniformly sample 32 frames from the video files to fit the maximum inputs of the chosen model: LLaVA-NeXT-Video-7B-hf. The reason for selecting this model is explained in detail in Section 4.1.3.

All 1,473 records are used in this research, but the signal and video data are not of the same duration for each record. We process the signal data to match the video data's time length. We demonstrate the data processing progress in Figure 2.

There are three tables for the signal data, vehicle data (VD), vehicle control data (VCD), and satellite data (SD) each. These data types were collected at different frequencies, resulting in varying

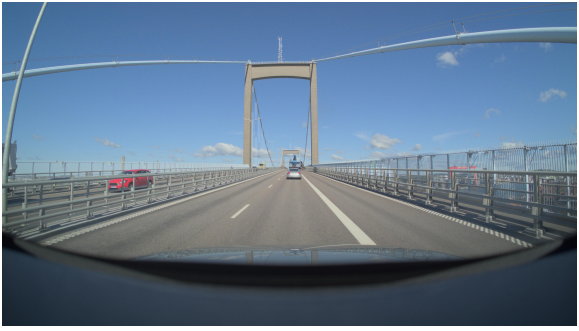

**Figure 1: A video frame from record #000005**

table lengths. To standardize the data, we downsized all tables based on the least frequent data, the satellite data. After reshaping, we have a table with approximately 1,700 rows. The signals (columns) that are difficult to interpret or remain constant are considered noise and, therefore, removed from the study.

## 3.2 Scenario Query

The scenario query processing pipeline (Figure 2) consists of two major steps. The first step is the textual description generation process, which converts the signal and video into textual information separately. The second step involves individually comparing the query text with the video and signal descriptions and then combining the results. We filter these results to retrieve the top records and, based on their IDs, locate the corresponding records in the database.

*3.2.1 Video Description Generation.* This paper aims to maximize the utility of video data by enabling the generation of detailed descriptions based on video content. The dataset facilitates the replication of our research and validation of the results. We selected three models for their performance and functionality, which are discussed in Section 4.1.3.

*3.2.2 Signal Description Generation.* Signal descriptions are created with the assistance of an industrial partner, using automated programs to convert signals into textual descriptions. These signals represent various properties of test drives, such as address, acceleration, and brake pedal input. Automated interpreters are preferred over LLMs because most properties have fixed values; interpreters provide faster and more accurate descriptions without relying on model-based learning.

*3.2.3 Query Processing.* We use a sentence similarity model, *all-MiniLM-L6-v2* to match queries to the natural language descriptions. This approach remains manageable and computationally cheap. The *all-MiniLM-L6-v2*[1] model, known for its efficiency, compares the cosine similarity between the embeddings of queries and record descriptions. This approach can help save storage space when the number of records are limited.

---

[1]https://huggingface.co/sentence-transformers/all-MiniLM-L6-v2

## 4 EXPERIMENTAL EVALUATIONS

### 4.1 Description Generation

*4.1.1 Prompts:* To answer RQ1, we designed six prompts (shown in Table 1) to generate descriptions using the LLaVa model. We focused on this model because it supports both single conversations (Prompt 1 - 4) and multi-run conversations (Prompt 5 - 6). Each prompt provides more information than the previous one, allowing us to assess how specific the prompt needs to be, to provide the relevant query results.

*4.1.2 Data Modalities:* To analyze how data modalities impact description generation, we use the same model (LLaVa) and prompts to generate descriptions for the same record using image and video data. This model is chosen for its ability to process both modalities. By comparing descriptions generated from a single video frame versus an entire video, we aim to identify differences and determine the one that provides clearer and more reliable information about the records.

*4.1.3 Models:* In this study, we examine three models for video description generation, as detailed in the following paragraphs. All three models, launched in 2024, can generate textual information from video.

*LLaVA-NeXT-Video-7B-hf.* (`LLaVa`) [23]: This model was fine-tuned using both image and video datasets. It was chosen for its ability to effectively interpret both static figures and dynamic videos, enabling us to compare the descriptions generated from each format and analyze the differences. Additionally, it supports multiple conversation rounds with the model, allowing us to compare various prompts.

*InternVideo2-Chat-8B.* (`InternVideo`) [18]: Built on the Mistral-7B architecture [6], this model integrates an LLM with a video-specific module (BLIP), creating a VideoLLM capable of understanding and responding to video content effectively.

*Cogvlm2-llama3-caption.* (`Cogvlm`) [19]: Based on LLaMa-3.1-8B [1], this model specializes in generating descriptive text for videos. It excels at capturing and clearly articulating motion features, particularly for analyzing dynamic content.

### 4.2 Scenarios Query

To verify the query results and examine how keywords affect the outcomes, we developed nine scenarios as listed in Table 2, ranging from simple to complex, with suggestions from developers at the collaborating company.

These scenarios were selected based on several factors. First, we considered the complexity and diversity of the scenarios to ensure they reflect a range of real-world applications. Second, feedback from developers helped us prioritize scenarios that are practical and relevant to actual usage. Third, we ensured that the chosen keywords were searchable within the dataset. Lastly, we aimed for the scenarios to cover both signal data (e.g., locations and addresses) and video data (e.g., weather and vehicles). Through these experiments, we address RQ2.

**Table 1: Prompts Used in the study and examples of generated descriptions for sample record #000005**

| ID | Prompt | Generated Descriptions for Record #000005 |
|----|--------|--------------------------------------------|
| 1 | Describe the driving conditions and environments that can be seen in the video. | The video shows a car driving on a bridge with a clear view of the road ahead. The car is on the right side of the road, and there is a white van in front of it. The bridge has a clear view of the road ahead, and the car is driving on the right side of the road. The video captures the car's journey on the bridge, with a clear view of the road ahead. |
| 2 | Given a video of the test drive, describe the driving conditions, environmental factors and other notable surrounding events and behaviors. | The video shows a car driving on a bridge with a clear sky and a few clouds in the background. The car is on the right side of the road, and there is a white van in front of it. [Repeated sentences] |
| 3 | Given a video taken from the dash camera of car during a test drive, generate a concise and detailed description of the scenario the car is in. Your output should provide a clear and comprehensive overview of the driving conditions, environmental factors, and any notable events or behaviors exhibited by the car during the test drive. Use only information gathered from the video. This information will be used by engineers to understand the scenario that the car is in. | The car is driving on a bridge with a clear and sunny day. The bridge is a suspension type with a single arch design. The car is on the right side of the bridge, and there are two lanes on the bridge. [Repeated sentences] |
| 4 | Interpret the video by identifying the current speed of the vehicle, acceleration, braking patterns, and general driving conditions. Also, analyze environmental factors like weather, lighting, and road conditions. Please describe any significant road events, such as vehicles overtaking, pedestrians crossing, or tunnels approaching. | The video shows a car driving on a bridge with a clear view of the road ahead. The car is moving at a moderate speed, and there are no visible pedestrians or tunnels approaching. The weather appears to be cloudy, and the lighting suggests it is daytime. The road conditions are clear, with no visible obstacles or debris. The car is driving on the right side of the road, and there are no vehicles overtaking or crossing in the immediate vicinity. The bridge is a suspension type, and the car is on the right side of the road. |
| 5 | Interpret the video by identifying the driving conditions, environmental factors and any significant road events. Driving conditions include: Identify and describe the speed of the vehicle; Detect any instances of acceleration or deceleration. Explain if braking occurs; If possible, estimate the force or intensity of the acceleration/braking. Environmental factors include: Analyze the weather (e.g., clear, rainy, snowy, foggy) and road conditions (e.g., wet, dry, icy). Significant road events include: Recognize and describe any other vehicles on the road, including whether they surpass the driver or stay ahead; Detect any pedestrians, cyclists, or other objects (e.g., road signs, traffic lights) in or near the road; Note any changes in the surroundings, such as tunnels, bridges, intersections, or sharp turns approaching; Report any traffic congestion, accidents, or emergency vehicles in view. Describe the video with the format of **Driving conditions: ... **; **Environmental factors: ... **;**Significant events: ... ** | **Driving conditions:** The video shows a car driving on a bridge with a clear sky and no visible traffic congestion. The car is moving at a moderate speed, and there are no indications of acceleration or deceleration. The road appears to be dry and well-maintained. **Environmental factors:** The weather is clear, and the sky is blue, suggesting a sunny day. The bridge is well-lit, and there are no visible signs of rain or other adverse weather conditions. **Significant events:** The car is driving on a bridge with no other vehicles in sight, and there are no pedestrians or cyclists visible. The road is clear, and there are no signs of accidents or emergency vehicles. The car is maintaining a steady speed and appears to be following the rules of the road. |
| 6 | Interpret the video by identifying the driving conditions, environmental factors, and any significant road events. **Driving Conditions:*** Identify and describe the speed of the vehicle.* Detect any instances of acceleration or deceleration. Explain if braking occurs.* If possible, estimate the force or intensity of the acceleration/braking.**Environmental Conditions:*** Analyze the weather (e.g., clear, rainy, snowy, foggy) and road conditions (e.g., wet, dry, icy).**Significant Road Events:*** Recognize and describe any other vehicles on the road, including whether they surpass the driver or stay ahead.* Detect any pedestrians, cyclists, or other objects (e.g., road signs, traffic lights) in or near the road.* Note any changes in the surroundings, such as tunnels, bridges, intersections, or sharp turns approaching.* Report any traffic congestion, accidents, or emergency vehicles in view. Describe the video with the format of **Driving conditions: ... **; **Environmental factors: ... **; **Significant events: ... ** | **Driving Conditions:** The video shows a car driving on a bridge with a clear sky and no visible traffic congestion. The car is moving at a moderate speed, and there are no signs of acceleration or deceleration. The driver maintains a steady pace throughout the video. **Environmental Factors:** The weather appears to be clear and dry, with no visible signs of rain, snow, or fog. The road is dry, and there are no indications of icy conditions. **Significant Events:** There are no other vehicles on the road, and no pedestrians or cyclists are visible. The car stays ahead of the camera's view throughout the video. The bridge is the main focus, and there are no tunnels, intersections, or sharp turns approaching. The video is a straightforward drive on a bridge with no significant events. |

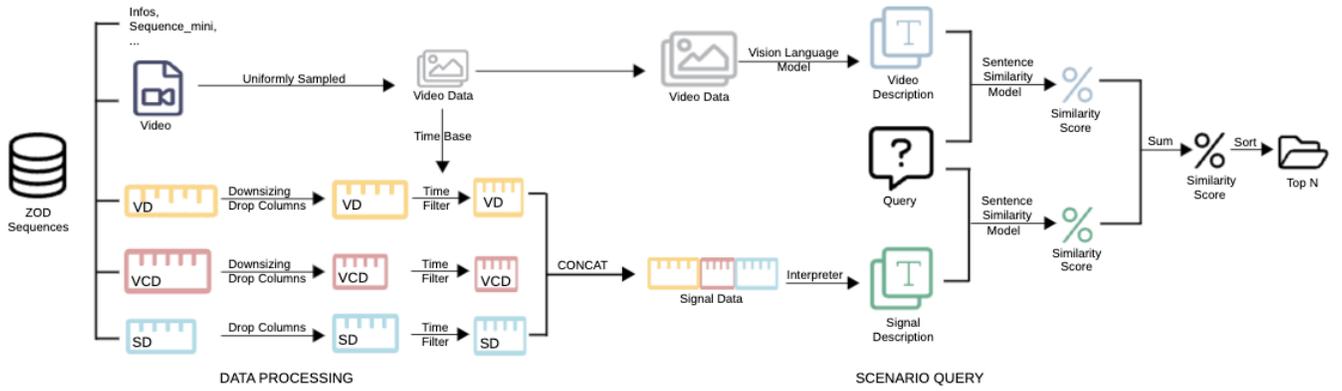

**Figure 2: The workflow contains two procedures: data processing and scenario query processing. When processing data, video data provides the time span for each record, and 32 uniformly selected frames are extracted. The signal data is down-sampled to match the lowest frequency record and time-filtered based on the video's time span. Finally, all three types of signal data are concatenated into a single table. After data processing, video and signal data are utilized to generate textual descriptions. The video data and textual descriptions are then independently compared to the query. The similarity scores from both comparisons are summed, and the results are ranked to retrieve the top N most similar records.**

**Table 2: Scenario Search Queries**

| ID | Scenario |
|---|---|
| I | Driving in the snow |
| II | Driving in the tunnel |
| III | Driving in France and pedestrians in the view. |
| IV | Driving on the bridge with a car ahead. |
| V | When driving in rain or snow and there is a car ahead of you, and you want to make a right turn. |
| VI | Driving on a cloudy day, approaching a bridge with no other cars in sight. |
| VII | The lane marking changes from a dashed line to a solid line when exiting a tunnel. |
| VIII | Driving on a highway, an exit ramp is ahead. |
| IX | Driving on a rural road in the afternoon, and the vertical acceleration in meters per second squared is greater than 9. |

## 4.3 Validation Metrics

One of the challenges in this work is evaluation, as the data lacks labels, making it difficult to determine whether we have found the best match for our query. However, during this study, we discovered that although direct evaluation is not feasible, we can still assess reliability by comparing the distance between the query and the matched record and the distances between the query and other records. We denote all records as being randomly labeled from 1 to 1473. Let $D_n$ represent the distance between the query and the record labeled $n$. We define the gap between consecutive distances as

$$G_n = D_{n+1} - D_n (1 \leq n \leq 1472),$$

To quantitatively evaluate the architectures and retrieval quality across different scenarios, we employed and evaluated the following metrics to answer RQ3:

- LGap (The Largest Gap): Represents the largest distance between two consecutive records, denoted as $LGap = \max(G_n)$. Larger gaps indicate low similarity between the generated descriptions.
- MinD (Min Distance): The distance between the query and the most similar record, reflecting how far the query deviates from the most similar generated description. Denoted as $MinD = \min(D_n)$.
- MaxD (Max Distance): The distance between the query and the least similar record, reflecting how far the query deviates from the least similar generated description. Denoted as $MaxD = \max(D_n)$.
- Range: The difference between MaxD and MinD indicates the range of similarity among the generated descriptions. A smaller difference suggests that most records are fairly similar to the query, while a larger difference indicates a greater diversity in how records relate to the query.
- StdDev (Standard Deviation): The standard deviation of distances across all records. A low standard deviation suggests most descriptions are similar, while a higher standard deviation indicates greater diversity among the generated descriptions.
- RLGap (Relative Largest Gap): Measures the largest gap in distance between consecutive records, normalized by the total range of distances. This provides a sense of how significant the largest gap is compared to the overall spread of the data. A higher Relative Largest Gap suggests significant differences in similarity among the records, indicating the possible correct answers.

When querying a specific scenario, the criteria for a reliable result includes:

- A few highly relevant records with similarity scores above 0.9, demonstrating the pipeline's ability to retrieve desired scenarios.

- Some irrelevant records should have similarity scores below 0.4, indicating the pipeline's effectiveness in filtering out unrelated responses.
- The remaining records, with similarity scores between 0.4 and 0.9, should be distributed without forming distinct clusters. This spread reflects the pipeline's ability to generate varied descriptions for different scenarios.

A key indicator of a failed search is low variance. When all records achieve similarly high or low scores, showing little distinction among descriptions.

To assess whether the query results follow this pattern, we can visualize them using a violin plot with box plot, highlighting the median and distribution of all records' distances from the query.

## 5 Results

### 5.1 RQ1

*5.1.1 Prompts:* To identify the effect of the prompts on the quality of generated descriptions. We compared the six different prompts, reviewed all generated text, and used record #000005 as an example in this paper. We present the results by showcasing a single video frame using Prompt 4 in Figure 3.

In Table 1, a multi-run-prompt consistently produced better descriptions than a single-line-prompt. Prompts 2 and 3 generated repetitive phrases, a pattern also observed in other instances. Starting from Prompt 4, the generated descriptions became more concise and clearer, without redundancy, and all information was directly inferred from the records. The descriptions produced by Prompt 5 and 6 showed no significant differences.

We also generated violin plots for each prompt, as illustrated in Figure 4. From the perspective of prompts:

- Prompt 4 generates the most reliable results. Its longer whiskers indicate a greater spread of data. Compared to other prompts, it has long needle-like shapes at both ends, suggesting some highly relevant (> 0.9) and less relevant (< 0.4) candidates. The lowest score is around 0.2 (Scenario VI), and the highest is close to 1 (Scenario IV). In the middle range (0.4–0.9), it demonstrates a well-distributed variance.
- Prompts 1, 2, and 3 struggle to identify highly relevant records (> 0.9) and fail to effectively filter out irrelevant ones (< 0.4). While they show good variance, they lack both high (> 0.9) and low (< 0.4) scores. Their median values are lower than those of Prompts 4, 5, and 6.
- Prompts 5 and 6 exhibit very low variance, with flat plots and no scores below 0.4, indicating that the generated descriptions are highly similar. Across all prompts, Prompt 5 and Prompt 6 exhibit higher median scores. This suggests that both prompts generate a large amount of information, making it more challenging to precisely match the query to the records.

Overall, multi-run prompts produced more concise and accurate descriptions than single-run prompts, while Prompt 4 offers a more cost-efficient option.

*5.1.2 Data Modality:* To assess the impact of data modality on the quality of generated descriptions, we showcased the description produced for a specific frame from the same video (#000005). Using prompt 4, the generated description is:

> From the perspective of the vehicle's speed, it seems to be maintaining a steady pace, which is appropriate for the conditions. The acceleration and braking patterns are not clearly visible, but the car appears to be maintaining a consistent speed without sudden acceleration or deceleration. The road ahead is clear, and there are no signs of heavy traffic or congestion, which indicates that the driver is likely following the speed limits and driving responsibly. The environmental factors include a cloudy sky, which might suggest overcast weather conditions, but no rain or snow is visible, and the lighting is bright, indicating it is daytime. The road is clear, with no visible debris or obstacles, which is good for driving conditions. The bridge's suspension design is evident, and the car is on.

However, this description includes details that cannot be inferred from a still image, such as "appears to be maintaining a consistent speed without sudden acceleration or deceleration," which requires motion to be observed.

We also reviewed all the descriptions generated in Prompt 6 and observed several phrases that never appear in image-based descriptions. For example, phrases like 'suggesting a slowing down' and 'approaching an intersection' demonstrate the ability to convey motion, which marks an improvement over image descriptions.

The video description shown in Table 1 is clearer, more concise, and easier to understand. Its lack of ambiguity, redundancy, and unnecessary detail makes it more suitable for efficient communication. Since the video description contains most of the important information in the image description, we do not consider adding the image description as one of the inputs.

*5.1.3 Model:* Based on our findings that Prompt 4 offers a more cost-efficient option and that video data provides additional information not obtainable from image data, we used Prompt 4 and video data to evaluate three selected models.

The results are shown in Figure 5. The plot generated by `LLaVa` (bottom) exhibits superior clustering. The long needle-shaped extension at the top reflects its ability to identify highly relevant matches (score > 0.9) while avoiding an overabundance of high-relevance responses, a drawback of `Cogvlm`. Additionally, `LLaVa` incorporates lower-scoring candidates (< 0.4), demonstrating its ability to filter out irrelevant responses—a weakness of `InternVideo`. The middle score range (0.4–0.9) remains well-balanced compared to `InternVideo`, preventing excessive high scores.

**RQ1:** Prompt 4 is a more cost-effective solution. Additionally, video data contributes valuable information that is inaccessible through image data. Among the models selected, the `LLaVa` model generated the most reliable descriptions, making it particularly suitable for querying.

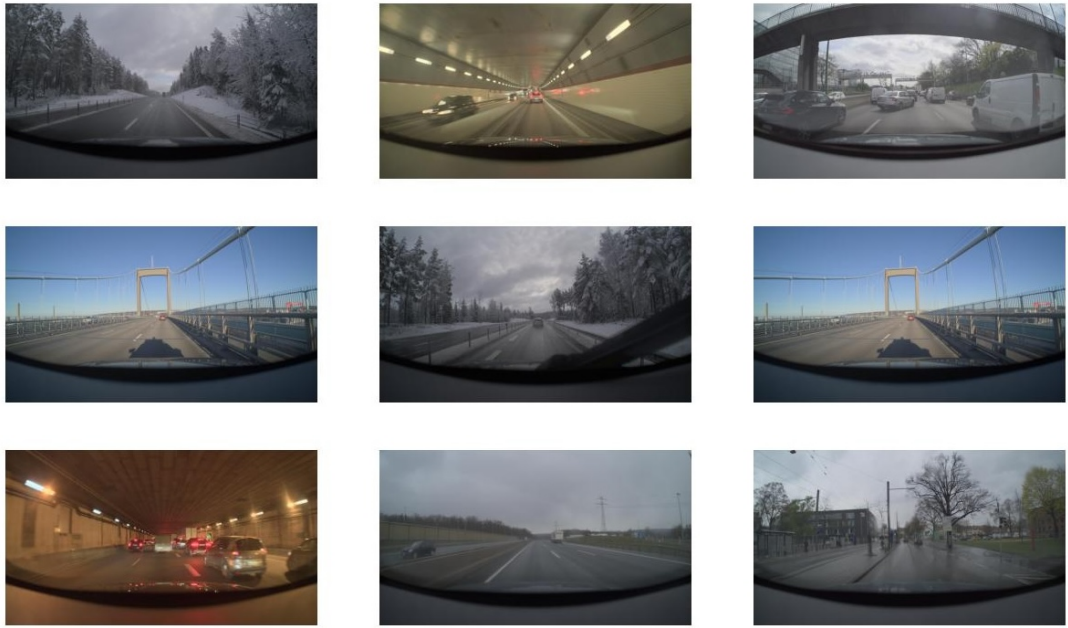

**Figure 3: Query results using Prompt 4 across scenario I to IX (from left to right, top to bottom)**

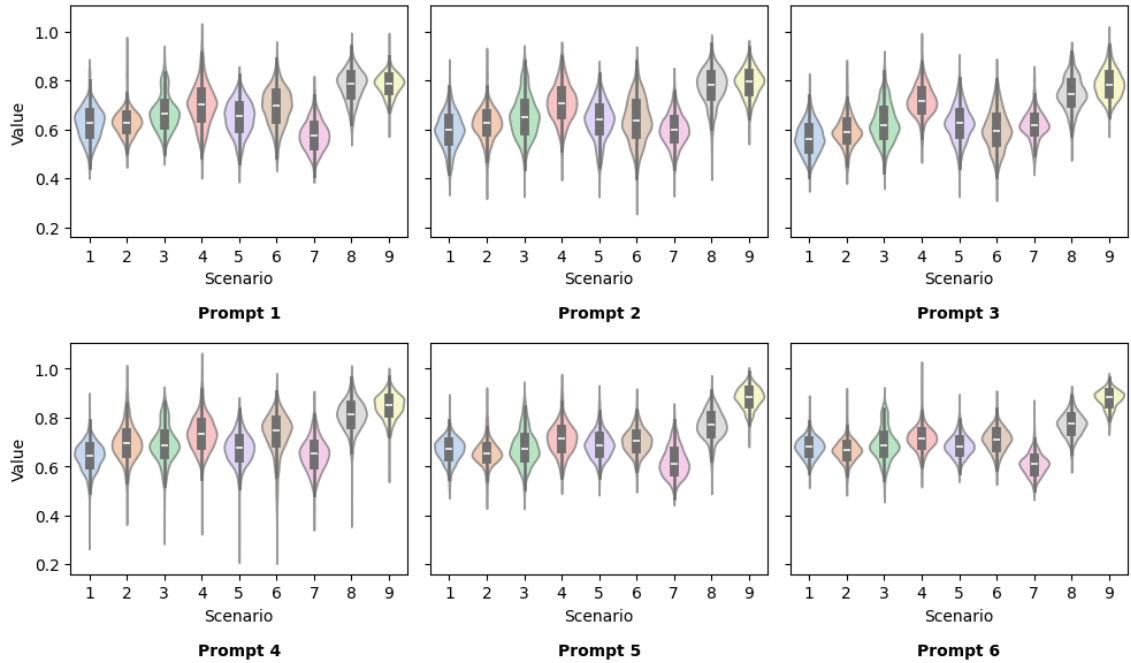

**Figure 4: Records' distances to nine scenarios' queries using six prompts. The Y-axis represents the cosine similarity score when calculating the distances between the embeddings of the query and the records' descriptions. The X-axis represents the Scenarios from I to IX.**

## 5.2 RQ2

Nine scenarios were designed from simple, single-keyword cases to more complex ones involving multiple keywords. As illustrated in violin plots in Figure 4, we categorize the values as follows:

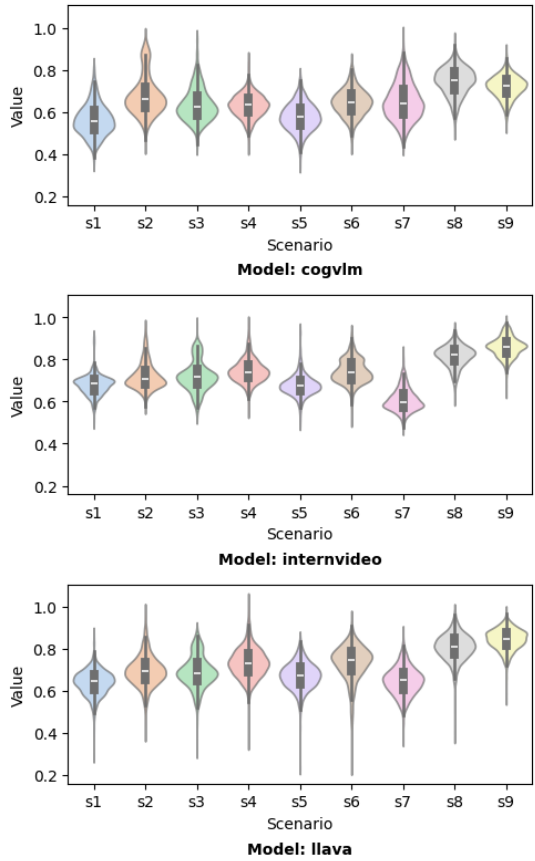

**Figure 5: All records' distances across three models. The Y-axis shows the distances between the search query and all records. The X-axis is Scenario I to IX.**

- Highly Relevant: The top narrow end. The top part of the violin, where values cluster above 0.9, indicates high similarity.
- Moderately Relevant: The wide middle section of the violin plot (Value 0.4 - 0.9). This range includes responses with a decent but imperfect similarity to the reference.
- Non-Relevant: The bottom narrow end. The bottom portion of the violin represents responses with low similarity score (< 0.4).

Scenario I and II are simple queries with only one keyword. Scenario I reaches a maximum of 0.9 across different prompts, with a median of 0.6. Prompts 4, 5, and 6 approach 0.7, while Prompt 3 falls below 0.6. Prompt 4 producing the best shape plot. Scenario II performs consistently well across all prompts, yielding highly relevant candidates (>0.9). Prompts 2 and 4 also capture some non-relevant responses (<0.4).

In Scenario III, the address "France" is inferred solely from signal data and accurately reflected in the descriptions. However, there are fewer highly relevant records compared to Scenario I and II.

Scenario IV performs well across all prompts, with the highest median and a distinct long needle shape at the top (>0.9). The middle range (0.4–0.9) is well-distributed with high variance. Scenario VI, which contains multiple keywords, encounters the terms cloudy, bridge, and cars more prominently observed in the video. It follows a similar pattern to Scenario IV, indicating that the bottleneck lies in feature detection rather than keyword count.

Scenario V contains more keywords than previous four scenarios, yet its median values across all prompts are lower. No records exceed 0.9 in similarity score. Scenario VII also includes more keywords, but the "change from a dashed to a solid line" is difficult to observe in the video. It has the lowest median score among all scenarios, with no highly relevant matches (>0.9).

Scenario VIII and Scenario IX maintain consistently high similarity scores across prompts, while others fluctuate more. Scenario IX has the most compact distribution, suggesting that generated descriptions are highly similar, making it difficult to identify strong matches. Rather than using multi-run conversation prompts, simpler prompts like prompts 3 and 4 might results in higher variance value and make it easier to identify the good matches.

Overall, when the query is short, but the descriptions are long, we noticed that the found records might not be so accurate. This is due to the increased presence of similar records and the long descriptions that can lower the similarity score, causing relevant records to be excluded from the results. Second, when the keywords are included in the descriptions but do not have the same semantic meaning, we found that the query results might not be accurate. For example, when the scenario III asks for pedestrians, it will find some records with descriptions of "no pedestrians."

We also developed an API to perform queries and visualize the results by displaying a video frame. This allows us to manually verify if the results match the queried scenarios. As mentioned, there are no established benchmarks for evaluating the results, a key challenge in this research. Using images helps verify visually identifiable features, such as snow or traffic lights, typically captured in the video data. However, some features, like speed or location, require manual checking of the signal data. We illustrate the Scenario I (Figure 6) and IX (Figure 7) as they are the best and worst case according to the curve analysis.

**RQ2:** Short queries combined with long descriptions will lower the similarity score, which can lead to the exclusion of relevant records from the retrieval results. Additionally, the results often lack accuracy when keywords are present in the descriptions but do not align semantically with the query. This highlights the importance of both query length and semantic consistency in optimizing data retrieval quality and effectiveness

## 5.3   RQ3

In addition to visualizing the records, we qualitatively analyze the results. The results of all experiments are summarized in Table 3. From the largest gap, we infer that as prompts become more detailed, the descriptions initially become more diverse but show diminishing differences after a certain point. In our experiments, most architectures exhibited the greatest diversity with prompt 4, which had the largest gap. This is likely linked to the length of the descriptions, as more detailed prompts tend to generate longer ones.

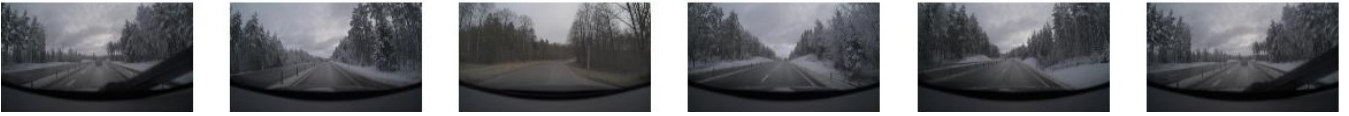

**Figure 6: Query results for scenario I using prompts 1 to 6 (from left to right)**

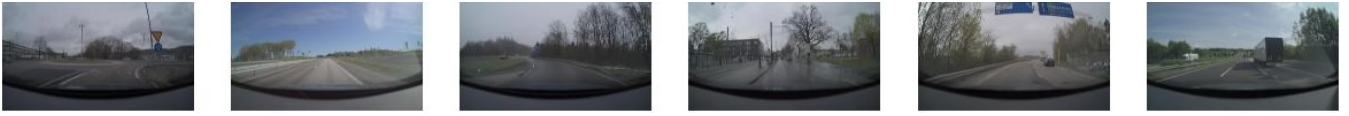

**Figure 7: Query results for scenario IX using prompts 1 to 6 (from left to right)**

Additionally, starting with prompt 4, a fixed format was enforced, possibly contributing to the lower gap scores and reduced diversity afterward.

We also calculated the maximum and minimum distances between the query and the records. Prompt 4 yielded both the highest and lowest distances. We further confirmed this by calculating the range, observing that Prompt 4 achieved the highest range across all scenarios. Regarding standard deviation, we found that in all scenarios, Prompts 5 and 6 had lower standard deviations than the previous prompts. For RLGap, the highest values occurred at different prompts. Prompt 4 achieved the highest value in four of the nine scenarios, indicating that all the generated descriptions exhibited better diversity and significant differences, a positive sign for accurate matching.

**RQ3:** Although there are no direct benchmarks for evaluating query results, reliable outcomes typically display consistent patterns. First, when plotting the sorted distances for all records, an 'S' shape typically emerges, helping to distinguish three categories: highly relevant, moderately relevant, and non-relevant items. Additionally, various metrics can be used to assess the quality of the results. A wide range of values and significant gaps between them generally indicate a successful match.

## 6   Discussion

Our study provides valuable insights for autonomous driving software companies, particularly in log storage optimization and retrieval for software testing. Test driving logs and associated data are primarily used for short-term development iterations with limited long-term utility. However, as simulation and cloud services evolve, more effective data retention and utilization strategies must be explored. Our research demonstrates that leveraging LLMs can significantly enhance log retrieval efficiency, reducing the workload of developers and enabling more effective use of historical log information. By integrating LLMs into log management systems, companies can streamline debugging, improve software validation, and accelerate development cycles.

**Limitation:** One limitation of this study is the publicly available and relatively small dataset. Given the limited scale, we could store and retrieve log descriptions through direct comparison without a significant computational overhead. However, as the scale of data grows in industrial applications, it will be important to implement a database for storing all descriptions to reduce computational costs. Additionally, once descriptions are generated, annotating and categorizing the data based on commonly used keywords should be considered to facilitate easier access for developers in real-world practice.

**Table 3: Comparison Of Metric Results Across Different Combinations of Scenarios And Prompts (show as S-P in the table)**

| S-P | LGap | MinD | MaxD | Range | StdDev | RLGap (%) |
|---|---|---|---|---|---|---|
| I-1 | 0.2813 | 0.4306 | .856 | 0.4254 | 0.067 | 66.14 |
| I-2 | 0.2706 | 0.3652 | 0.8531 | 0.4879 | 0.0718 | 55.45 |
| I-3 | .282 | .376 | .7998 | 0.4237 | 0.0631 | 66.56 |
| I-4 | 0.4428 | 0.2888 | .874 | 0.5852 | 0.0581 | 75.66 |
| I-5 | 0.2319 | 0.4925 | .872 | 0.3795 | 0.0488 | 61.11 |
| I-6 | 0.2027 | 0.5322 | .8708 | 0.3386 | 0.0416 | 59.87 |
| II-1 | 0.3598 | 0.4724 | 0.9511 | 0.4788 | 0.059 | 75.16 |
| II-2 | 0.3064 | 0.3483 | 0.9018 | 0.5534 | 0.0687 | 55.36 |
| II-3 | 0.3722 | 0.4084 | 0.8548 | 0.4464 | 0.063 | 83.38 |
| II-4 | 0.3537 | 0.3956 | 0.9805 | 0.5849 | 0.0728 | 60.47 |
| II-5 | 0.2452 | 0.4504 | 0.8997 | 0.4493 | 0.05 | 54.59 |
| II-6 | 0.2416 | 0.5028 | 0.8999 | 0.3971 | 0.0444 | 60.84 |
| III-1 | 0.2319 | 0.491 | 0.9087 | 0.4176 | 0.0723 | 55.52 |
| III-2 | 0.3547 | 0.3651 | 0.9042 | 0.5391 | 0.0863 | 65.80 |
| III-3 | 0.281 | 0.394 | 0.8851 | 0.4912 | 0.0777 | 57.21 |
| III-4 | 0.4083 | 0.3173 | 0.8918 | 0.5744 | 0.0747 | 71.07 |
| III-5 | 0.2458 | 0.4594 | 0.9138 | 0.4545 | 0.0708 | 54.08 |
| III-6 | 0.329 | 0.4853 | 0.8931 | 0.4078 | 0.0669 | 80.68 |
| IV-1 | 0.3734 | 0.4399 | 0.9943 | 0.5543 | 0.0852 | 67.36 |
| IV-2 | 0.3053 | 0.4285 | 0.9236 | 0.4951 | 0.0764 | 61.67 |
| IV-3 | 0.3375 | 0.4944 | 0.9643 | 0.4699 | 0.0638 | 71.81 |
| IV-4 | 0.4548 | 0.3547 | **1.0317** | 0.677 | 0.0702 | 67.18 |
| IV-5 | 0.288 | 0.5182 | 0.9488 | 0.4306 | 0.064 | 66.88 |
| IV-6 | 0.3077 | 0.5387 | 1.0056 | 0.4669 | 0.0492 | 65.91 |
| V-1 | 0.2671 | 0.4174 | 0.8261 | 0.4086 | 0.0694 | 65.37 |
| V-2 | 0.3984 | 0.3598 | 0.8442 | 0.4843 | 0.077 | 82.26 |
| V-3 | 0.3101 | 0.3583 | 0.8746 | 0.5162 | 0.0717 | 60.06 |
| V-4 | 0.5073 | **0.2345** | 0.8541 | 0.6195 | 0.0618 | 81.89 |
| V-5 | 0.2214 | 0.5074 | 0.9071 | 0.3996 | 0.0538 | 55.40 |
| V-6 | 0.2012 | 0.557 | 0.876 | 0.319 | 0.0431 | 63.08 |
| VI-1 | 0.3208 | 0.4655 | 0.9245 | 0.4589 | 0.0761 | 69.90 |
| VI-2 | 0.3377 | 0.2958 | 0.8971 | 0.6013 | **0.0907** | 56.16 |
| VI-3 | 0.3747 | 0.3454 | 0.8521 | 0.5067 | 0.0798 | 73.94 |
| VI-4 | **0.5515** | 0.2411 | **0.9432** | **0.7021** | 0.083 | 78.55 |
| VI-5 | 0.2494 | 0.5188 | 0.8944 | 0.3756 | 0.0514 | 66.39 |
| VI-6 | 0.2065 | 0.5481 | 0.8876 | 0.3395 | 0.0482 | 60.82 |
| VII-1 | 0.3222 | 0.4114 | 0.7904 | 0.3789 | 0.0611 | **85.02** |
| VII-2 | 0.3125 | 0.3568 | 0.8225 | 0.4657 | 0.0636 | 67.11 |
| VII-3 | 0.3009 | 0.4405 | 0.8326 | 0.3921 | 0.0569 | 76.76 |
| VII-4 | 0.2751 | 0.3697 | 0.8781 | 0.5084 | 0.066 | 54.11 |
| VII-5 | 0.2346 | 0.4699 | 0.8293 | 0.3594 | 0.0611 | 65.28 |
| VII-6 | 0.1988 | 0.4844 | 0.8509 | 0.3665 | 0.0447 | 54.23 |
| VIII-1 | 0.2684 | 0.5655 | 0.9662 | 0.4006 | 0.0628 | 67.00 |
| VIII-2 | 0.386 | 0.4282 | 0.9547 | 0.5264 | 0.0726 | 73.33 |
| VIII-3 | 0.265 | 0.5049 | 0.9269 | 0.422 | 0.0676 | 62.80 |
| VIII-4 | 0.4945 | 0.382 | 0.9846 | 0.6025 | 0.0617 | 82.07 |
| VIII-5 | 0.2628 | 0.5143 | 0.9465 | 0.4322 | 0.0573 | 60.82 |
| VIII-6 | 0.2017 | 0.5982 | 0.9072 | 0.309 | 0.0479 | 65.27 |
| IX-1 | 0.2181 | 0.592 | 0.9741 | 0.3821 | 0.0449 | 57.09 |
| IX-2 | 0.2303 | 0.5663 | 0.9396 | 0.3734 | 0.056 | 61.68 |
| IX-3 | 0.221 | 0.597 | 0.9946 | 0.3976 | 0.0585 | 55.58 |
| IX-4 | 0.3086 | 0.559 | 0.9804 | 0.4215 | 0.0489 | 73.22 |
| IX-5 | 0.1573 | 0.7002 | 0.986 | 0.2858 | 0.0432 | 55.04 |
| IX-6 | 0.1495 | 0.7476 | 0.9646 | 0.217 | **0.0372** | 68.89 |

Another limitation is the availability of signal data in the dataset. In practice, the number of signal data points is significantly larger than that publicly accessible in open datasets. In real-world applications, to maximize the utility of signal data, it is necessary to categorize them into discrete and continuous values. By leveraging LLMs, we can analyze continuous values more effectively and generate descriptions that better capture trends over time. Developing robust methods to process and interpret these signal data will be crucial for improving the performance and applicability of log retrieval systems.

**Future Work:** In future research, we aim to improve the dynamic description of signal data, particularly for continuously changing signals, to better capture trends and patterns over time. Additionally, we plan to enhance the architecture of the query engine by exploring more advanced sentence similarity models. Due to resource constraints, we could not train a customized model for this study; however, access to larger and more diverse open datasets could enable the development of tailored models with improved performance. Exploring these advancements will further refine the potential of LLMs in autonomous driving software testing and log management.

## 7 Threads to Validity

A threat to **internal validity** arises in selecting prompts and scenario-based search queries. We mitigate this by using a broader range of prompts and queries and incorporating real-world scenarios from industry practice. As a threat to **external validity**, evaluating the proposed pipeline against one open dataset may limit generalizability. Nonetheless, the selected dataset is the only publicly available dataset with paired video and signal records from our knowledge. We minimize this threat by relying on the expertise of our collaborating industry partner. A threat to **conclusion validity** stems from data sourced from a single company and collected only within European countries. We address this by selecting scenarios based on diverse criteria to enhance representativeness.

## 8 Conclusions

In this work, we explored the potential of combining video data with signal data to enhance data querying and improve log retrieval in autonomous driving applications. We studied factors affecting description generation and found that video data produces more trustworthy and concise descriptions than image data alone. While multi-run prompts can generate more detailed descriptions, they may reduce query accuracy, as overly long descriptions can be less effective when dealing with short queries. A more cost-efficient approach, such as Prompt 4 in our study, provides a balanced trade-off between detail and accuracy. Among the models evaluated, LLaVa performed the best, demonstrating strong capabilities in identifying highly relevant cases while effectively filtering out non-relevant ones.

We identified challenges in detecting keywords related to signal data or motion-related features, which can lead to false positives, through nine evaluated scenarios. In contrast, keywords associated with video data were easier to detect, though the model occasionally struggled with semantic confusion and the misinterpretation of negation. To support practical implementation, we developed an API that visualizes query results, integrating curve plots as indicators of search quality alongside various metrics to assess retrieval accuracy and overall effectiveness.

Since no established metrics exist for evaluating this problem, we introduced range, standard deviation, and relative gap to quantitatively measure results. These proved to be effective evaluation metrics. Additionally, violin plots combined with box plots offer a clear visualization of query reliability. Integrating these visualizations with video frames enables developers to better assess and validate the scenarios they are searching for.

Our findings highlight the advantages of integrating video and signal data for more accurate and efficient log retrieval in autonomous driving applications. By identifying key challenges in description generation and query accuracy, we provide insights into optimizing prompt strategies and model selection, with LLaVa emerging as the most effective option. Additionally, our proposed evaluation metrics and visualization techniques offer a practical framework for improving search reliability, aiding developers in scenario analysis and system refinement. Future research can build upon our approach by refining retrieval models, incorporating larger datasets, and exploring advanced methods for better-integrating signal and video data.

## Acknowledgments

We would like to thank Zenseact for their supports in the data analysis.

Software Center has also partially funded this work, a collaboration between the University of Gothenburg, Chalmers, and 18 universities and companies – www.software-center.se.

ChatGPT was used to improve spelling, grammar, punctuation, and clarity.